\documentclass[aps,prl,twocolumn,showpacs,preprintnumbers,amsmath,amssymb]{revtex4-2}

 \def\frac#1#2{{#1\over #2}}

\def\be{\begin{equation}}
\def\ee{\end{equation}}
\newcommand{\bea}{\begin{eqnarray}}
\newcommand{\eea}{\end{eqnarray}}
\newcommand{\ket}[1]{\left| #1 \right>}
\newcommand{\bra}[1]{\left< #1 \right|}

\usepackage{color}
\usepackage{graphicx}
\usepackage{dcolumn}
\usepackage{bm}
\usepackage{hyperref}

\begin{document}

\title{Quantum complexity and topological phases of matter}

\author{Pawel Caputa and Sinong Liu}

\affiliation{Faculty of Physics, University of Warsaw, ul. Pasteura 5, 02-093 Warsaw, Poland. }

\date{\today}

\begin{abstract}
In this work, we find that the complexity of quantum many-body states, defined as a spread in the Krylov basis, may serve as a new probe that distinguishes topological phases of matter. We illustrate this analytically in one of the representative examples, the Su-Schrieffer-Heeger model, finding that spread complexity becomes constant in the topological phase. Moreover, in the same setup, we analyze exactly solvable quench protocols where the evolution of the spread complexity shows distinct dynamical features depending on the topological vs non-topological phase of the initial state as well as the quench Hamiltonian. 
 
\end{abstract}

\maketitle

\section*{Introduction}
Recently, the problem of quantifying quantum complexity attracts a lot of attention in different branches of physics. To a large extent, these developments are stimulated by the holographic correspondence \cite{Ma} and puzzles related to the nature of black holes in quantum gravity. Among others, complexity plays essential roles in discussions on the firewall paradox \cite{Harlow:2013tf}, fast scrambling near black hole horizons \cite{Susskind,Aaronson,Stanford:2014jda,Brown:2015bva}, quantum chaos \cite{Maldacena:2015waa}, and finally the nature of the holographic dictionary itself \cite{Bouland:2019pvu}. 

What makes holography powerful, is the dual description of quantum gravity in terms of a strongly interacting quantum system. 
In that framework, rather than just stating that a certain state is simple or complex (e.g. assigning it to one of the complexity classes) one is interested in a precise measure of state's complexity that can be computed on both sides of the holographic correspondence. This demand sparked a lot of new developments focused on complexity in more general many-body systems and quantum field theories (QFTs) (see e.g. \cite{ComplexQFT} and reviews \cite{Chapman:2021jbh,Chen:2021lnq} for complete list of references). Now, as the dust is starting to settle, we are provided with new, potentially very powerful, tools for exploring the complexity frontier of field theories. 

One of the important directions where these complexity measures can be tested and may provide a new looking glass, is the physics of topological phases of matter \cite{Hasan:2010xy,Qi:2010qag,Schnyderetal}. Indeed, the importance of complexity in topological order was realised early on (see e.g. \cite{DoritA,MilMiy}). On the other hand, applying geometric ideas from holography to topological phases proved fruitful in \cite{Wen:2016bud,Aguado:2007oza,Matsuura:2010vi,Czech:2022wtt}, so clearly these two subjects may have a lot to learn from each other. Nevertheless, the actual studies of state complexity measures in models with topological phases are in their infancy. 

So far, only the Nielsen-type approach to complexity \cite{Nielsen1}, was applied to a couple of examples \cite{Ali:2018aon,Liu:2019aji,Xiong:2019xoh}. Since this framework, tailor-made for computer science, gives rise to numerous ambiguities in many-body physics, such as the choice of gates with their penalties, cost functions etc., it is not surprising that conclusions on whether complexity is sensitive enough differed between \cite{Ali:2018aon} and \cite{Liu:2019aji,Xiong:2019xoh}. Still \cite{Liu:2019aji,Xiong:2019xoh} argued that, with an appropriate choice for the ambiguities, one may detect different phases by discontinuities in Nielsen's complexity. Based on the above, whether or not complexity is a universal probe correlated with topological phases remains an open question.

In this work we shed new light on this issue by analyzing a recent measure of state complexity called {\em spread complexity} \cite{Balasubramanian:2022tpr} in one of the simplest models with topological phases, the Su-Schrieffer-Heeger (SSH) model \cite{Su:1979ua}. Spread complexity admits a clear definition valid in arbitrary quantum systems (including free and interacting field theories), is relatively straightforward to compute and already proved handy in diagnosing quantum chaos \cite{Balasubramanian:2022tpr}. A closely related K-complexity \cite{Parker:2018yvk} has also been extensively studied as a good probe of the operator growth in many-body systems. \\  We first analytically compute the spread complexity of formation of the ground state and find that it is sensitive to the two phases of the model. More precisely, it becomes constant in the topological phase.
Next, we test its evolution following quantum quench protocols and find that it also distinguishes the phases even in this non-equilibrium scenario. Our analytical results not only reinforce the spread complexity as a powerful tool but also give a hope for universality in studying topological phases from the perspective of complexity.
\section{Spread Complexity of Quantum States}
We begin with a brief review of the spread complexity \cite{Balasubramanian:2022tpr} and the Krylov basis that plays the main role in the computations. Further details can be found in  \cite{Balasubramanian:2022tpr,Caputa:2021sib}.\\ We will be interested in complexity of a general quantum state $\ket{\Psi(s)}$ related to some initial $\ket{\Psi_0}$ by a unitary transformation
\be
\ket{\Psi(s)}=e^{-iHs}\ket{\Psi_0}.\label{Circs}
\ee
In the quantum computation parlance,  $\ket{\Psi(s)}$ can be referred to as a ``target state" related to a ``reference state" $\ket{\Psi_0}$ by a ``unitary circuit" with a ``circuit Hamiltonian" $H$. Parameter $s$, usually $s\in[0,1]$, denotes a ``circuit time", can be also taken arbitrary and regarded as the physical time $t$ (see below).\\
A useful measure of quantum state complexity can be defined by a way that some initial state $\ket{\Psi_0}$ is spread in the Hilbert space by a unitary $U(s)$ \cite{Balasubramanian:2022tpr}. Intuitively, a complex ``evolution" will lead to a fast spread over all orthogonal states. More precisely,  the spread complexity of  $\ket{\Psi(s)}=U(s)\ket{\Psi_0}$ is estimated by the minimum over all choices of basis $\mathcal{B}=\left\{\ket{B_n}, n=0,1,2,...| \ket{B_0}=\ket{\Psi_0} \right\}$ of the following cost function
\be
\mathcal{C}(s)=\min_{\mathcal{B}}\left(\sum_n n|\langle \Psi(s)|B_n\rangle|^2\right).\label{ComplGen}
\ee
The fact that makes this definition powerful and computable is that the minimum is attained (see \cite{Balasubramanian:2022tpr} for proofs) when basis $\mathcal{B}$ is the so-called Krylov basis.

The idea behind the Krylov basis for state \eqref{Circs} is to consider states with all the different powers of the circuit Hamiltonian acting on the initial state $\ket{\Psi_0}$, i.e., $\{\ket{\Psi_0},H\ket{\Psi_0},H^2\ket{\Psi_0},...\}$, and apply the Gram–Schmidt orthogonalization procedure, known as the Lanczos algorithm \cite{LanczosVish}, to this set. In this new basis $\ket{K_n}$, the circuit Hamiltonian $H$ is generally tri-diagonal and acts as 
\be
H\ket{K_n}=a_n\ket{K_n}+b_n\ket{K_{n-1}}+b_{n+1}\ket{K_{n+1}},
\ee
where coefficients $a_n$ and $b_n$ are the so-called Lanczos coefficients. The information about them is also contained in the moments of the return-amplitude (auto-correlator) $S(s)\equiv\langle\Psi(s)|\Psi_0\rangle$. 

Having constructed the basis that minimizes \eqref{ComplGen}, we expand our state as
\be
\ket{\Psi(s)}=\sum_n\psi_n(s)\ket{K_n},\label{StateKB}
\ee
where, by construction, the complex coefficients in this expression satisfy a discrete Schrodinger equation
\be
i\partial_s\psi_n(s)=a_n\psi_n(s)+b_{n}\psi_{n-1}(s)+b_{n+1}\psi_{n+1}(s).\label{Schreq}
\ee
With the knowledge of the Lanczos coefficients, we can solve it with initial condition $\psi_{n}(0)=\delta_{n,0}$ (so that we start from $\ket{K_0}=\ket{\Psi_0}$), and determine \eqref{StateKB}. Note that unitarity implies that $\sum_n(p_n(s)\equiv|\psi_n(s)|^2)=1$. Moreover, the return amplitude is related to the first coefficient by $S(s)=\psi^*_0(s)$. Last but not the least, the number of the independent Krylov vectors depends on the Hamiltonian as well as the initial state $\ket{\Psi_0}$.\\
Most importantly, in the Krylov basis, the complexity \eqref{ComplGen} becomes
\be
\mathcal{C}(s)=\sum_n n|\psi_n(s)|^2,\label{ComplKrB}
\ee
and for all practical purposes, this will be our working definition of the spread complexity in the remaining part of this paper. This measure naturally generalizes the Krylov complexity (K-complexity) of operators \cite{Parker:2018yvk} to quantum states. Recent studies indicate that this new notion of complexity can distinguish integrable and chaotic models \cite{Parker:2018yvk,Rabinovici:2022beu}.  Moreover, the evolution of the so-called thermofield-double state \cite{Takahashi:1996zn} leads to return amplitude given by the spectral form factor (see e.g. \cite{Guhr:1997ve,Cotler:2016fpe}) making spread complexity a new probe of quantum chaos. Even though we are only starting to explore this universal new tool it is clear that its sensitivity to interesting physics is tantalising \cite{Trigueros:2021rwj,Dymarsky:2021bjq,Caputa:2021ori,Yates:2021lrt}.  In the following, we will employ \eqref{ComplKrB} and test wether it does equally well in integrable systems and in particular, whether it can detect topological phases. 
\section{Spread complexity in the SSH model}
Our basic example will be the SSH model of polyacetylene \cite{Su:1979ua} (see e.g. \cite{TopInS} for a pedagogical introduction, here we closely follow the conventions of \cite{RH,Miyaji:2014mca}) given by the Hamiltonian 
\bea
H&=& t_1 \sum_i \left( c_{Ai}^{\dagger} c_{Bi} + \text{h.c.} \right) -  t_2 \sum_i \left( c^{\dagger}_{Bi} c_{A,i+1} + \text{h.c.} \right)\nonumber\\
&+& \mu_s \sum_i \left( c^{\dagger}_{Ai} c_{Ai} - c^{\dagger}_{Bi} c_{Bi} \right),\label{HamiSSH}
\eea
where $(c_{Ai}, c_{Bi})$ represent two-flavours of fermion annihilation operators defined at site $i$ on a 1d lattice, $t_1, t_2, \mu_s$ are real parameters, and we also assume anti-periodic boundary conditions (see more in appendix A). We will take both $t_1,t_2 \ge 0$  and $\mu_s=0$. Depending on these couplings, the model is in one of the two phases: a non-topological phase for $t_1>t_2$ or a topological phase (topological insulator) for $t_2>t_1$, separated by a critical point at $t_1=t_2$. 

We will first compute the complexity of the ground state that, depending on parameters $t_1$ and $t_2$, belongs to one of the above-mentioned phases of the model. For that, as well as for later purposes, it will be convenient to re-write the Hamiltonian in momentum space as (see \cite{Miyaji:2014mca}, and appendix A) 
\be
H=\sum_{k}\left[2R_3 J^{(k)}_0+iR_1\left(J^{(k)}_+-J^{(k)}_-\right)\right],
\ee
where the coefficients are $R_1=t_1 - t_2 \cos (k)$, $R_3= t_2 \sin (k)$ and, for each momentum mode, we denoted the $SU(2)$ algebra generators 
\be
[J^{(k)}_0,J^{(k)}_\pm]=\pm J^{(k)}_\pm,\qquad [J^{(k)}_+,J^{(k)}_-]=2J^{(k)}_0.
\ee
\begin{figure}[t!]
\centering
\includegraphics[width=0.4\textwidth]{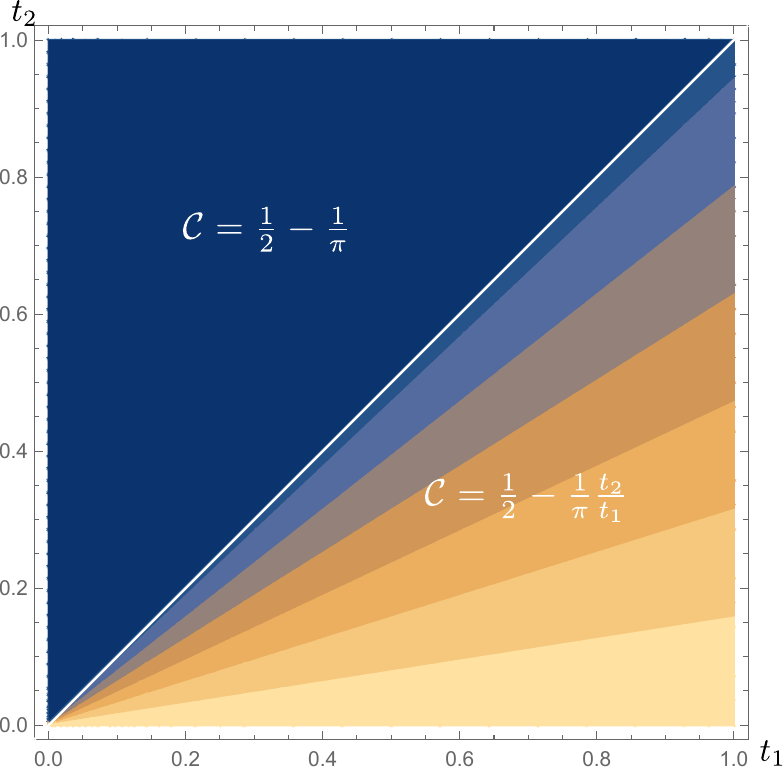}
\caption{Behaviour of the spread complexity of formation $\mathcal{C}(t_1, t_2)$ (equation \eqref{kcmpl1_rs}) for the ground state of the SSH model $|\Omega \rangle$.}
\label{kcmpl1}
\end{figure}
Then, the ground state can be written in terms of the $SU(2)$ coherent states as (see appendix A)
\be
\ket{\Omega}=\prod_{k>0}\mathcal{N}_ke^{-i\tan\left(\frac{\phi_k}{2}\right)(J^{(k)}_++J^{(-k)}_+)}\ket{\frac{1}{2},-\frac{1}{2}}_k,
\ee
where $\mathcal{N}_k$ stands for the normalization, $J^{(\pm k)}$ correspond to two decoupled $SU(2)$ algebras for positive and negative momenta and $\ket{1/2,-1/2}$ denotes a tensor product over the lowest-weight states of the $j=1/2$ representation ($J^{(\pm k)}_{-}\ket{1/2,-1/2}_{\pm k}=0$).  Moreover, the relation between $\phi_k$ and the physical parameters is given by
\be
\sin\phi_k=\frac{|R_1|}{R},\qquad \cos\phi_k=\frac{R_3}{R},\label{ParamPhi}
\ee
where we also denoted 
\be
R=\sqrt{t_1^2 + t_2^2 - 2t_1 t_2 \cos (k)}.\label{RGen}
\ee
Without loss of generality, we can just compute the spread complexity for positive momenta and the full result will have an additional factor of 2 from the $-k$ sector. 

First, for a single momentum $k>0$, using the $SU(2)$ Baker–Campbell–Hausdorff formula and with a slight abuse of notation, we write the relevant part of the state in a circuit form \eqref{Circs}
\be
\ket{\Omega_k(s)}=e^{-i\frac{s\phi_k }{2}\left(J^{(k)}_++J^{(k)}_-\right)}\ket{\frac{1}{2},-\frac{1}{2}}_k,
\ee
where $s\in[0,1]$ and our ground state is the ``target state" at $s=1$. The operator in the exponent is the circuit Hamiltonian in  \eqref{Circs}. Note that in these circuits we took a natural reference state $\ket{\Psi_0}$ as the ground state of the left and right Hamiltonians (see appendix A and also \cite{PCSL} for other choices). This way of writing makes transparent the connection with coherent states and we can directly apply the tools from \cite{Caputa:2021sib} (see appendix B) to expand our state in the Krylov basis as \eqref{StateKB}. Because $j=1/2$, we will only have two basis vectors and two amplitudes
\be
\psi_0(s)=\cos\left(\frac{s\phi_k}{2}\right),\qquad \psi_1(s)=-i\sin\left(\frac{s\phi_k}{2}\right),
\ee 
that satisfy \eqref{Schreq} with appropriate Lanczos coefficients. As a result, we get the contribution to our complexity from a single momentum mode
\be
\mathcal{C}_k(s=1)=\sin^2\frac{\phi_k}{2}=\frac{1}{2}- \frac{t_2 \sin (k)}{2\sqrt{t_1^2 + t_2^2 - 2t_1 t_2 \cos (k)}}. \
\ee
The complexity of the ground state is obtained by integrating over all the momenta and multiplying by 2 from $k<0$. This yields
\bea
\mathcal{C}(t_1,t_2) = 2\int_0^{\pi} \frac{dk}{2\pi} \mathcal{C}_k= \frac{1}{2} - \frac{t_1+t_2 -|t_1-t_2| }{2\pi t_1} 
\label{kcmpl1_rs}.
\eea
Observe that we took the continuum limit so this result is proportional to the volume $L$ but, to keep our equations compact, we rescaled this factor (our $\mathcal{C}$ are complexity ``densities").

This surprisingly simple formula, shown on Fig.\,\ref{kcmpl1}, indeed shows two very different behaviours of the spread complexity for the two distinct phases of the model. Namely, for the non-topological phase with $t_1>t_2$, complexity linearly depends on the ratio $t_2/t_1$ but in the topological phase, with $t_2>t_1$, it is constant. This is our main result. We also performed analogous computation for the 1d Kitaev-chain \cite{Kitaev:2000nmw}, and found that complexity becomes constant when crossing from a non-topological to a topological phase in specific cases \cite{PCSL} (see supplementary material D). Note that unlike entanglement entropy or Nielsen-type complexities that require numerics, \eqref{kcmpl1_rs} is fully analytical. 
\section{Complexity during quantum quench}
Another framework where we can probe the spread complexity is given by the so-called quantum quenches. A typical quench protocol considers a unitary time evolution of an initial state $\ket{\Omega_i}$ of some initial Hamiltonian $H_i$, performed with a different Hamiltonian $H_f$ for which $\ket{\Omega_i}$ is an excited state. Universal features of the evolution of entanglement and complexity have been extensively studied in the literature (see e.g. \cite{Mitra,Calabrese:2016xau,Camargo:2018eof,Liu:2019qyx} and review of closely related dynamical quantum phase transitions \cite{Heyl:2017blm}). 

Here, we focus on the so-called instantaneous quench in the SSH model and consider the state 
\begin{equation}
|\Psi (t) \rangle = e^{-i H_f t}| \Omega_i\rangle, \label{QuenchSSH}
\end{equation}
where $| \Omega_i\rangle$ is taken as the ground state of the initial Hamiltonian $H_{i}$ with parameters $(t^{i}_1,t^i_2)$, and the evolution is performed with the SSH Hamiltonian $H_f$ with different parameters $(t^{f}_1,t^f_2)$.
For this state, we will be interested in the Krylov basis and universal features of the real-time evolution of the spread complexity defined above. 

Curiously, the Nielsen-type geometric complexity measures were analyzed during such quenches \eqref{QuenchSSH} in the SSH model before \cite{Ali:2018aon}. In this work, authors argued that the above-mentioned complexities were neither sensitive to distinguish topological phases nor to whether the evolution is driven by $H_f$ or $H_i$. This gives us another strong motivation to compare with the spread complexity and its sensitivity in the out-of-equilibrium applications.

Similarly to the ground state, since the SSH Hamiltonian consists of the $SU(2)$ generators (for each momentum mode), we can employ coherent state techniques to expand the quench state \eqref{QuenchSSH} in the Krylov basis and find contribution to spread complexity from each mode. This analysis is based on the return amplitude (also known as Loschmidt amplitude) that for each mode $k>0$ reads
\be
S_k(t)=\bra{\Psi(t)}\Psi(0)\rangle_k=\cos(R_ft)-i\cos(\phi_f-\phi_i)\sin(R_f t). \label{RAQ}
\ee
In this formula, we used the notation $R_{i/f}$ that simply stands for \eqref{RGen} with parameters $(t^{i/f}_1,t^{i/f}_2)$ and similarly $\phi_{i/f}$ are related to the momentum and parameters as in \eqref{ParamPhi} with $(t^{i/f}_{1},t^{i/f}_{2})$. From \eqref{RAQ} we can repeat the whole procedure to extract Lanczos coefficients, derive $\psi_n(t)$ and compute spread complexity as before. The contribution from a single momentum mode becomes 
\be
\mathcal{C}_k(t)=\sin^2(\phi_f-\phi_i)\sin^2(R_f t).\label{CQPerM}
\ee
Clearly, for each mode, complexity is periodic in time with periods governed by the parameters of the evolving Hamiltonian $H_f$. The spread complexity is then not symmetric under the change of parameters of initial and final Hamiltonians. The amplitude depends on both, initial and final parameters.
\begin{figure}[t!]
\centering
\includegraphics[width=0.45\textwidth]{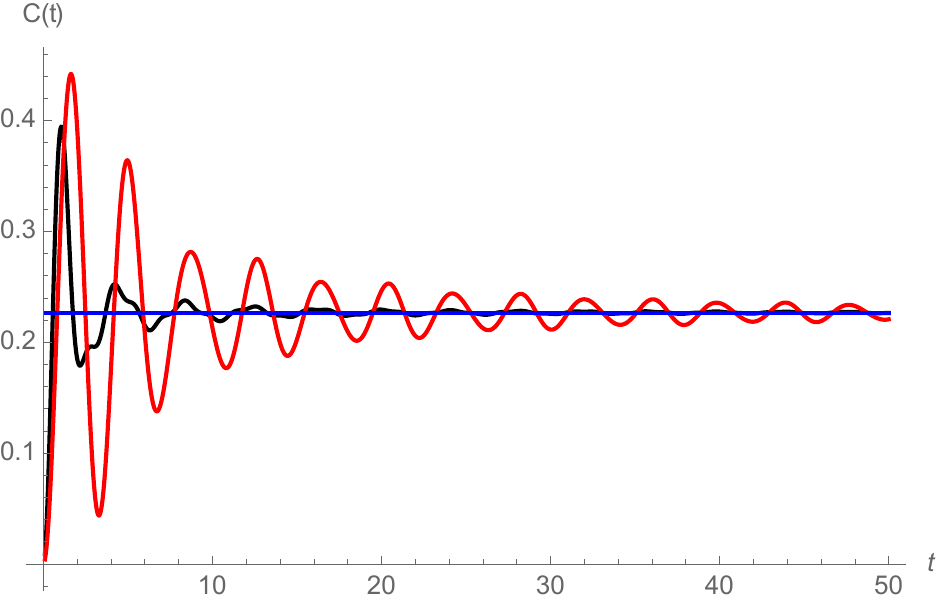}
\caption{Evolution of spread complexity starting from $H_i$ with $(t^i_1=1,t^i_2=0.2)$ and $(t^f_1=0.7,t^f_2=1.5)$ (in black) and the same parameters in $H_i$ and $H_f$ switched (in red). At late times complexity approaches constant \eqref{LTCon} (in blue).}
\label{kcmpl2}
\end{figure}

The total complexity is obtained by integrating over the momenta and multiplying by the factor of 2 from negative $k$. In the continuum, this can be written in a compact form as an integral (again a density)
\begin{equation}
\mathcal{C}(t) =2 \int_0^{\pi} \frac{dk}{2\pi} \frac{\left( t^f_{1} t^i_{2} - t^i_{1} t^f_{2} \right)^2 \sin^2 (k) \sin^2 \left( R_f t \right) }{R^2_f R^2_i},\label{ComEvolGen}
\end{equation}
and for general parameters we need to integrate it numerically. Let us first study some of the analytical examples. Clearly, the spread complexity vanishes if $t^{i}_1=t^{f}_1=0$ or $t^{i}_2=t^{f}_2=0$. Hence, this measure assigns zero costs to evolution of states with Hamiltonians on the axes of the $t_1-t_2$ plane, as long as they are evolved with Hamiltonian on the same axis (that may have a different parameter $t^{i/f}_2$ in the first or $t^{i/f}_1$ in the second family).\\
A more interesting analytical example is derived if one of the parameters of $H_f$ vanishes i.e., $t^f_{1(2)}=0$. The total spread complexity is then oscillating with frequency $t^f_{2(1)}$
\be
\mathcal{C}(t) =\frac{\left(t^i_1+t^i_2-|t^i_1-t^i_2|\right)^2}{8(t^i_{2(1)})^2}\sin^2\left(t\,t^f_{2(1)}\right),
\ee
and the amplitude of these oscillations distinguishes the two phases of the initial quench state $\ket{\Omega_i}$.

Generally, note that the integral would be symmetric under the change of $i$ and $f$ labels if not the $\sin^2(R_f t)$ part. We can then pick two Hamiltonians, each with a pair of generic parameters in each of the two phases, and compare the evolution of the ground state $\ket{\Omega_i}$ of $H_i$ in the topological phase with $H_f$ in the non-topological phase, and other way around (by just switching parameters $t^i_a$ with $t^f_a$ in \eqref{ComEvolGen}). This is plotted in Fig. \ref{kcmpl2} and, unlike Nielsen complexities studied  in \cite{Ali:2018aon}, we can verify that our complexity measure can distinguish the two different quench protocols \cite{CommentSH}.  This is our second key finding. 

In addition, we can derive universal results analytically for the early as well as late time evolution of the spread complexity. For that, it is useful to parametrize $t^i_2\equiv\alpha\, t^i_1$ and $t^f_2\equiv\beta\, t^f_1$, such that the topological and non-topological phases are separated by $\alpha,\beta=1$. Then, in the early times we have a universal quadratic growth 
\be
\mathcal{C}(t)\sim (\alpha-\beta)^2\frac{(1+\alpha-|1-\alpha|)^2}{8\alpha^2}(t^f_1)^2 t^2,
\ee
so this early time regime distinguishes whether $\ket{\Omega_i}$ is in the topological or non-topological phase. \\
Also for late times, as it can be observed from Fig. \ref{kcmpl2}, complexity approaches a constant that is given by 
\bea
&&\mathcal{C}(t\to\infty)=\frac{(\alpha -\beta )}{8 \alpha  \beta }\times\nonumber\\
&& \left(\beta-\alpha +\frac{\alpha  (\beta +1) | \beta -1|
   -\beta(\alpha +1)  | \alpha -1|}{1-\alpha  \beta }\right).\label{LTCon}
\eea
Clearly, this constant only depends on the two combinations, $\alpha$ and $\beta$, of the four parameters of the model and is symmetric under their exchange. Therefore, it is the same for both protocols (see Fig. \ref{kcmpl2}) discussed above but it depends on the phases of the model. Note that this constant is the effect of summing of all the, otherwise periodic, contributions from different momenta. We have analyzed it in more detail in appendix C and it is clear that this {\em spread complexity constant} can be also used as a probe of the late-time physics of the quench state. 
\section{Conclusions and discussion}
In this letter, we showed that the spread complexity based on the Krylov basis \cite{Balasubramanian:2022tpr} is a sensitive probe correlated with topological phases. We found that, for the ground state of the SSH model, the complexity depends on the ratio of parameters $t_1/t_2$ in the trivial phase, whereas it becomes constant in the topological insulator regime. Based on related analysis in the Kitaev model, we expect that this behaviour is universal and further tests (including numerics) in more general, interacting models will be very important to verify this observation. 

For the quantum quench, our main goal was to demonstrate the advantage of the spread complexity over more geometric measures studied in \cite{Ali:2018aon}. In addition, we found various universal features of the evolution, e.g. at early and late times, that are sensitive to the topological phases. Certainly, further analysis of the dynamics of spread complexity (e.g. during slow or fast-type quenches, with time-dependent Hamiltonians or Floquet driving) is a promising future direction. Along the same lines, it will be very interesting to study the spread complexity (and generally Krylov complexity tools) in the context of dynamical quantum phase transitions \cite{Heyl:2017blm}. 

Finally, it will be very important to understand the relation between the spread complexity and other topological order parameters already studied in the literature, such as Berry phases or Chern number, that can even be extracted experimentally using certain quench protocols \cite{Tarnowski}. These links may play an important role in sharpening the spread complexity as a tool in condensed matter and many-body physics.




\section{Appendix A: Details of the SSH model}\label{AppendixA}
In this appendix we provide more technical details of the SSH model [26] and few steps to express it in the language of the $SU(2)$ coherent states. This material should be sufficient to understand our formulas in the main text.\\
First, via the Fourier transform (We assumed anti-periodic boundary conditions such that the momenta in the first Brillouin zone take values $k_n = \frac{2 \pi}{L}\left(  -\left\lceil \frac{L-1}{2} \right\rceil  +  n + \frac{1}{2}  \right) ,  n = 0, 1, 2, \cdots, L-1$. Below we neglect the index $n$.)  
\be
\begin{pmatrix}
c_{Al} \\
c_{Bl} \\
\end{pmatrix}
 = N^{-1/2} \sum_{k \in \text{BZ}} e^{ikl} 
\begin{pmatrix}
\tilde{c}_k \\
\tilde{d}_k \\
\end{pmatrix},
\ee
and a rotation of Pauli matrices $(\sigma_x, \sigma_y, \sigma_z)$ $\to$ $(\sigma_x, \sigma_z, - \sigma_y)$, the Hamiltonian in position space (equation (7) in the main text) turns into a free-fermion Hamiltonian in momentum space
\be
H = \sum_{k \in \text{BZ}} \Psi^{\dagger}_k \vec{R}(k) \cdot \vec{\sigma} \Psi_k, \quad 
\Psi_k =  
\begin{pmatrix}
c_k \\
d_k \\
\end{pmatrix}=e^{-\frac{i\pi}{4}\sigma_x}\begin{pmatrix}
\tilde{c}_k \\
\tilde{d}_k \\
\end{pmatrix},
\label{HamiSSHM}
\ee
where the vector of parameters reads
\be
\nonumber
\vec{R}(k) =  \begin{pmatrix}
R_1 \\
R_2 \\
R_3 \\
\end{pmatrix}=
\begin{pmatrix}
t_1 - t_2 \cos k \\
-\mu_s \\
t_2 \sin k \\
\end{pmatrix}.
\ee
The Hamiltonian \eqref{HamiSSHM} can be further diagonalised as follows:
\be
H = \sum_{k \in \text{BZ}} \chi_k ^{\dagger} \operatorname{diag} (R(k) , -R(k)) \chi_k,  
\ee
where we wrote the norm of the three-vector $R(k) = \left| \vec{R} (k) \right|$, and
\be
\chi_k =
\begin{pmatrix}
\chi_{+,k} \\
\chi_{-.k} \\
\end{pmatrix} = 
\begin{pmatrix}
\vec{v}_+(k) & \vec{v}_-(k) \\
\end{pmatrix}^{\dagger} \Psi_k.
\label{BogoT}
\ee
In this formula, we also introduced orthonormal eigenvectors $ \vec{v}_{\pm}(k)$ as
\be
\vec{v}_{\pm}(k) = \frac{1}{\sqrt{2R(R\mp R_3)}}
\begin{pmatrix}
R_1 - i R_2 \\
\pm R -R_3 \\
\end{pmatrix}.
\label{evpm}
\ee
In this notation, the ground state of SSH Hamiltonian can be written as
\be
|\Omega \rangle = \prod_k \chi_{-,k}^{\dagger} |0 \rangle
\label{gsSSH}
\ee
where state $|0 \rangle$ is the Fock vacuum of the free-fermion Hamiltonian \eqref{HamiSSHM}.

On the other hand, \eqref{HamiSSHM} can be decomposed into three parts -- analogues of left- and right-moving continuum Dirac Hamiltonians
\be
H_L =   \sum_k R_3 c^{\dagger}_{k} c_{k},  \qquad
H_R =  - \sum_k R_3 d^{\dagger}_{k} d_{k}, \label{HamiD}
\ee
as well as the mass part
\be
H_{LR} =  \sum_k \Psi^{\dagger}_k \left( R_1 \sigma_1 + R_2 \sigma_2 \right) \Psi_k.
\ee
The ground states of $H_L$ and $H_R$, denoted as $|G_L \rangle$ and $|G_R\rangle$, respectively, are given by
\be
| G_L \rangle = \prod_{k<0} c^{\dagger}_{k} |0 \rangle_L, \qquad  
| G_R \rangle = \prod_{k>0} d^{\dagger}_{k} |0 \rangle_R,
\ee
where $|0 \rangle_{L,R}$ is the Fock vacuum of $H_{L,R}$ respectively.\\
It is not hard to see that \eqref{BogoT} is a Bogoliubov transformation between the creation and annihliation operators of $H_L + H_R$ in \eqref{HamiD} and $H$ in \eqref{HamiSSHM}. Therefore the (unnormalized) ground state $|\Omega \rangle$ defined in \eqref{gsSSH} can be expressed in terms of $|G_L \rangle$ and $|G_R\rangle$ as
\be
\ket{\Omega}=\exp\left[-\sum_{k>0}\left(\frac{v_k}{u_k}c^\dagger_kd_k+\frac{u_{-k}}{v_{-k}}d^\dagger_{-k}c_{-k}\right)\right]\ket{G_L}\otimes\ket{G_R},
\ee
where
\be
\begin{pmatrix}
u_k \\
v_k \\
\end{pmatrix}
= \frac{1}{\sqrt{2R (R+R_3)}} 
\begin{pmatrix}
R+ R_3 \\
R_1 - i R_2 \\
\end{pmatrix}.
\ee
This is also the form presented in [39].

Let us now rewrite this state using the coherent state language. First, we introduce the $\operatorname{SU}(2)$ algebra generators $J^{(k)}_0$, $J^{(k)}_\pm$ (and similarly $J^{(-k)}_0$, $J^{(-k)}_\pm$) in terms of the fermionic creation and anihilation operators (from eq. (2) above)
\begin{eqnarray}
 J^{(k)}_+=-ic^\dagger_k d_k, &\quad& J^{(k)}_-=id^\dagger_k c_k, \nonumber \\
 J^{(k)}_0 &=& \frac{1}{2}(c^\dagger_k c_k-d^\dagger_k d_k), \\
 J^{(-k)}_+=-id^\dagger_{-k} c_{-k}, &\quad& J^{(-k)}_-=i c^\dagger_{-k} d_{-k}, \nonumber \\
J^{(-k)}_0 &=& -\frac{1}{2}(c^\dagger_{-k} c_{-k}-d^\dagger_{-k} d_{-k}),
\end{eqnarray}
where $k >0$.
Thus, the ground state \eqref{gsSSH} can be rewritten as a product of generalised $\operatorname{SU}(2)$ coherent states (for each momentum $k$)
\bea
| \Omega \rangle &=& \prod_{k>0} \cos^2 \frac{\phi_k}{2}  e^{ - i \tan \frac{\phi_k}{2} \left( e^{-i \psi_k} J_+^{(k)}  +  e^{i \psi_k} J_+ ^{(-k)} \right) }\nonumber\\
&& \qquad \qquad  \bigotimes_{k \in \text{BZ}} \left| 1/2; -1/2 \right\rangle_k. 
\eea
This is the form that we quoted in the main text.\\
Finally, the Hamiltonian of the SSH model can be rewritten in terms of $\operatorname{SU}(2)$ algebra generators as 
\bea
H =\sum_{k>0} R && \left[ 2 \cos \phi_k  \left(J_0^{(k)} + J_0^{(-k)} \right)  \right. \nonumber \\
&& \quad  +i \sin \phi_k \left( e^{-i \psi_k}  J_+^{(k)} +e^{i \psi_k} J_+^{(-k)} \right)   \nonumber \\
&& \quad \left.  -i \sin \phi_k \left( e^{i \psi_k}  J_-^{(k)} +e^{i \psi_k} J_-^{(-k)}\right)  \right],
\eea
where we used spherical coordinates $(\phi_k,\psi_k)$ (do not confuse $\psi_k$ with coefficients in the Krylov basis in the main text) to parametrise the components of $\vec{R}(k)$ as
\bea
\sin \phi_k \cos \psi_k &=& \frac{R_1}{R}, \nonumber \\
 \sin \phi_k \sin \psi_k &=& \frac{R_2}{R},  \nonumber  \\
\cos \phi_k &=& \frac{R_3}{R}.
\eea
In the main text we have worked with $\mu_s=0$ that corresponds to $R_2=\psi_k=0$ in these expressions.
\section{Appendix B: SU(2) coherent states}\label{Appendix B}
In this appendix we also collect some of the tools from the $SU(2)$ coherent states that were used in the main text (see more in e.g. [50,51]). We start with $su(2)$ algebra $[J_i,J_j]=i\epsilon_{ijk}J_k$, expressed in terms of the ladder operators $J_{\pm}=J_1\pm i J_2$ and $J_0=J_3$ as
\be
[J_{+},J_{-}]=2J_0,\qquad [J_0,J_\pm]=\pm J_\pm.
\ee
The generalised coherent states are defined by action of a displacement operator on the lowest-weight state $\ket{j,-j}$ (i.e., $J_-\ket{j,-j}=0$) labeled by (half or) integer $j$
\be
\ket{z,j}=e^{\xi J_+-\bar{\xi} J_-}\ket{j,-j}=e^{zJ_+}e^{\ln(1+z\bar{z})J_0}e^{-\bar{z}J_-}\ket{j,-j}.
\label{gnBCH}
\ee
In the above, we defined a complex parameter
\be
\xi=\frac{\theta}{2}e^{-i\varphi},\qquad 0\le\theta<\pi,\quad 0\le\varphi\le 2\pi,
\ee
with its complex conjugate $\bar{\xi}$ as well as
\be
z=\tan\left(\frac{\theta}{2}\right)e^{-i\varphi},\qquad \bar{z}=(z)^{*}
\ee

These coherent states are employed in the Krylov/spread complexity context where we e.g. chose $\xi=-is \frac{\phi}{2}$, so that $\theta=s\phi$ and $\varphi=\pi/2$. This way, the coherent state describing the unitary evolution in the Krylov basis is
\bea
\ket{z,j}=e^{-is\frac{\phi}{2}(J_++J_-)}\ket{j,-j}=\sum^{2j}_{n=0}\psi_n(s)\ket{K_n},
\eea
where the coefficients (amplitudes) are given by
\be
\psi_n(s)=\frac{(-i)^n\tan^n\left(\frac{s\phi}{2}\right)}{\left(\cos\frac{s\phi}{2}\right)^{-2j}}\sqrt{\frac{\Gamma(2j+1)}{n!\Gamma(2j-n+1)}},\label{AmplSU2}
\ee
and Krylov basis vectors are
\be
\ket{K_n}=\ket{j,-j+n}=\sqrt{\frac{\Gamma(2j-n+1)}{n!\Gamma(2j+1)}}J^n_{+}\ket{j,-j}.\label{KBV}
\ee
The amplitude \eqref{AmplSU2} satisfies a discrete Schrodinger equation ((5) in the main text) with $a_n=0$ and $2j+1$ coefficients
\be
b_n=\frac{\phi}{2}\sqrt{n(2j-n+1)},\qquad n=0,...,2j.
\ee
The Krylov or spread complexity in this case becomes
\be
\mathcal{C}(s)=\sum_n n|\psi_n(s)|^2=2j\sin^2\left(\frac{s\phi}{2}\right).
\ee
In the main text, we only needed a subset of these results for the $j=1/2$ representation where
\be
\mathcal{C}(s)=|\psi_1(s)|^2=\sin^2\left(\frac{s\phi}{2}\right).
\ee
More generally, we may consider time evolution 
\be
\ket{\Psi(t)}=e^{-iHt}\ket{j,-j},
\ee
with $su(2)$ coherence preserving Hamiltonian
\be
H=\gamma J_0+\alpha(J_++J_-).\label{HmasUGen}
\ee
Then the coherent state can be written as
\be
\ket{\Psi(t)}=\sum^{2j}_{n=0}\psi_n(t)\ket{K_n},
\ee
with Krylov basis vectors \eqref{KBV} and coefficients
\be
\psi_n(t)=e^{-jB}A^n\sqrt{\frac{\Gamma(2j+1)}{n!\Gamma(2j-n+1)}},\label{KoefPsit}
\ee
where
\bea
A&=&\frac{1}{i\sqrt{1+\frac{\gamma^2}{4\alpha^2}}\cot\left(\alpha t\sqrt{1+\frac{\gamma^2}{4\alpha^2}}\right)-\frac{\gamma}{2\alpha}},\\
e^{-jB}&=&\left(\cos\left(\alpha t\sqrt{1+\frac{\gamma^2}{4\alpha^2}}\right)+\right.\nonumber\\
&&\left.\frac{i\gamma}{2\alpha\sqrt{1+\frac{\gamma^2}{4\alpha^2}}}\sin\left(\alpha t\sqrt{1+\frac{\gamma^2}{4\alpha^2}}\right)\right)^{2j}.
\eea
Coefficients $\psi_n(t)$ in \eqref{KoefPsit} solve the Schrodinger equation (5) with Lanczos coefficients
\be
a_n=\gamma(-j+n),\qquad b_n=\alpha\sqrt{n(2j-n+1)}.
\ee
Finally, the spread complexity in this more general scenario becomes
\be
\mathcal{C}(t)=\sum_{n}n|\psi_n(t)|^2=\frac{2j}{1+\frac{\gamma^2}{4\alpha^2}}\sin^2\left(\alpha t\sqrt{1+\frac{\gamma^2}{4\alpha^2}}\right).
\ee
For the quench computation, we also needed the following return amplitude
\be
S(t)=\bra{j,-j}D^\dagger(z_i)e^{iH_ft}D(z_i)\ket{j,-j},
\ee
where the $SU(2)$ displacement operator is given by
\be
D(z_i)=e^{-i\frac{\phi_i}{2}(J_++J_-)},
\ee
and a Hamiltonian $H_f$ has a form \eqref{HmasUGen} with $\gamma=2R_3$ and $\alpha=iR_1$.\\
Generally, using the BCH relation we can evaluate this return amplitude explicitly 
\be
S(t)=\left(\cos(R_ft)-i\cos(\phi_f-\phi_i)\sin(R_f t)\right)^{2j}.
\ee
For $j=1/2$, its moments are
\be
\mu_n=R^n_f\left(\cos\frac{\pi n}{2}-i\cos(\phi_f-\phi_i)\sin\frac{\pi n}{2}\right),
\ee
and they correspond to the Lanczos coefficients
\bea
a_0&=&-R_f\cos(\phi_f-\phi_i),\qquad a_1=R_f\cos(\phi_f-\phi_i),\nonumber\\
b_1&=&R_f|\sin(\phi_f-\phi_i)|,
\eea
while all the higher ones vanishing. For our quench computations, these techniques again allowed us to map our state expanded in the Krylov basis to $SU(2)$ coherent state as
\bea
\ket{\Psi(t)}=e^{-iH_t t}\ket{\Omega_i}=
\sum^1_{n=0}\psi_n(t)\ket{K_n},
\eea
with amplitudes
\be
\psi_0(t)=S(t)^*,\quad \psi_1(t)=-i|\sin(\phi_f-\phi_i)|\sin(R_f t),
\ee
and the two basis vectors $\ket{K_0}=\ket{1/2,-1/2}$, $\ket{K_1}=\ket{1/2,1/2}$.
These formulas give rise to the spread complexity per mode given by equation (15) in the mian text.\\

Finally, observe that for $j=1/2$ representations of $SU(2)$, since we only have two complex amplitudes $\psi_0(t)$ and $\psi_1(t)$, we can relate the spread complexity $\mathcal{C}(t)=|\psi_1(t)|^2$ to the return amplitude $S(t)=\bar{\psi}_0(t)$. Indeed, by unitarity, probabilities sum up to identity and we have
\be
|\psi_0|^2+|\psi_1|^2=1\leftrightarrow\mathcal{C}(t)=1-|S(t)|^2.
\ee
This simple relation will not hold for a more general time evolution that requires higher number of Krylov basis vectors. Nevertheless, understanding the relation between spread complexity and Loschmidt echo is an interesting future problem. 
\section{Appendix C: Early and late times}\label{Appendix C}
This appendix contains a few more details on the universal features of the evolution of complexity in the early and late times.
\begin{figure}[t!]
\centering
\includegraphics[width=0.44\textwidth]{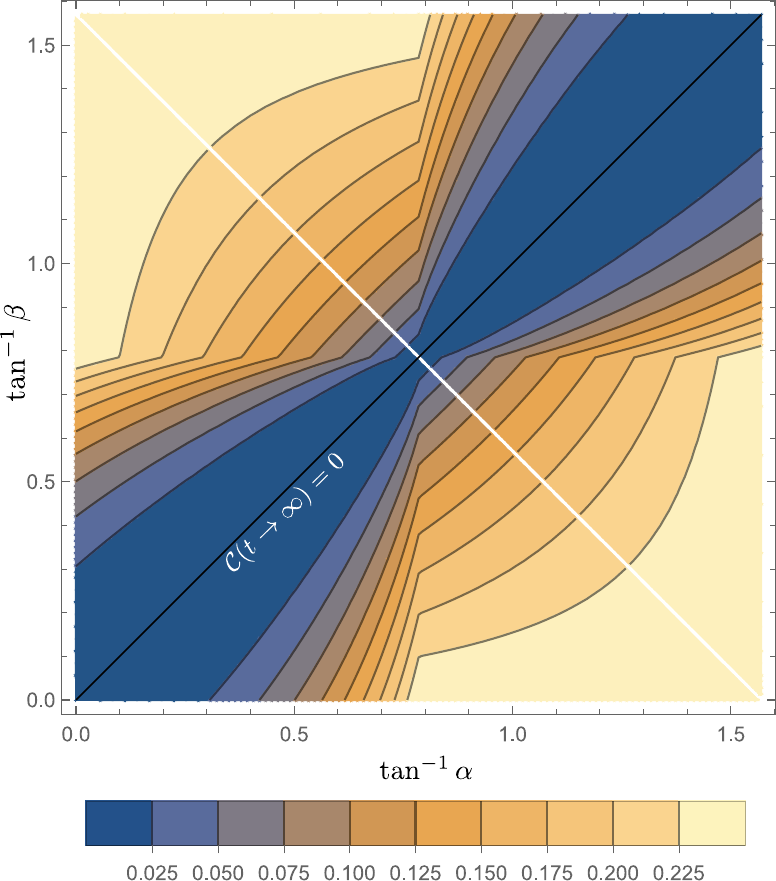}
\caption{Spread complexity constant at late times $\mathcal{C}(t \to \infty ; \alpha , \beta)$, shown in (23) in the main text.}
\label{kc_qq}
\end{figure}
Firstly, at the early times, after of the quench, i.e., $t \sim 0^+$, the time evolution of complexity (20) becomes
\be
\mathcal{C}(t \sim 0) \simeq t^2 \times \int_0^{\pi} \frac{dk}{\pi} \frac{\left( t^f_{1} t^i_{2} - t^i_{1} t^f_{2} \right)^2 \sin^2 (k) }{ R^2_i}.
\ee
In terms of the ratios of $t_2^{i,f}/t_1^{i,f}$, denoted as $\alpha$ and $\beta$ for initial and final parameters, respectively, the spread complexity is
\bea
\mathcal{C}(t \sim0) &=& (t\, t^f_{1})^2 \left( \alpha -\beta \right)^2   \frac{\left(|1+ \alpha| - |1-\alpha|\right)^2  }{8 \alpha^2}\nonumber \\
&=&  t^2 \left( t^f_{1} t^i_{2} - t^i_{1} t^f_{2} \right)^2\frac{\left( t_1^i + t_2^i - \left| t_1^i - t_2^i \right| \right)^2}{8 \left(t_1^i  t_2^i \right)^2}.\nonumber\\
\eea
On the other hand, for late times, i.e., $t \to \infty$, we notice that the time evolution of (20) can be estimated as follows. First, our formula is equivalent to
\be
\mathcal{C}(t) =2 \int_0^{\pi} \frac{dk}{2\pi} \frac{\left( t^f_{1} t^i_{2} - t^i_{1} t^f_{2} \right)^2 \sin^2 (k) }{R^2_f R^2_i} \times \frac{1- \cos 2R_f t}{2}.
\ee
We can then change the integration variable in the second term such that
\be
\begin{split}
 & \int_0^{\pi}  \frac{dk}{2\pi} \frac{\left( t^f_{1} t^i_{2} - t^i_{1} t^f_{2} \right)^2 \sin^2 (k) }{R^2_f R^2_i}  \cos (2R_f t)  \\
& =\frac{\left( t^f_{1} t^i_{2} - t^i_{1} t^f_{2} \right)^2}{t_1^f t_2^f} \int_{k(R_f) \in (0,\pi)}\frac{dR_f}{2\pi} \frac{ \sin k(R_f) }{R_f \cdot R^2_i (R_f)} \cos (2R_f t).
\end{split}
\ee
Now, because for $t \to \infty$ the cosine is highly oscillating in the interval $(0, |t_{1,f} \pm t_{2,f}|]$, where the integral range belongs to, this integral is always negligible, no matter what the concrete form of $\frac{ \sin k(R_f) }{R_f \cdot R^2_i (R_f)}$ in terms of $R_f$ is. This way we find the late time constant as
\be
\mathcal{C}(t \to \infty) \simeq \int_0^{\pi} \frac{dk}{2\pi} \frac{\left( t^f_{1} t^i_{2} - t^i_{1} t^f_{2} \right)^2 \sin^2 (k) }{R^2_f R^2_i} .
\ee
Above estimate is also equivalent to replacing the time dependent factor $\sin^2(R_ft)$ by $1/2$. This expression integrates to (23) in the main text and is plotted in Fig. \ref{kc_qq}. Note that naively, (23) is singular for $\alpha\beta=1$ (the white diagonal on Fig. \ref{kc_qq}), $\alpha=0$ and $\beta=0$. These cases should be analyzed more carefully starting from (20) with these values of $\alpha$ and $\beta$ and only then preforming the integral over momenta.

\section{Appendix D: One-Dimensional Kitaev Model}\label{Appendix D}
In this appendix we provide one more example where we can compute spread complexity using our general formalism. More details will appear in longer publication [41].\\
Our example is the 1d Kitaev model described by Hamiltonian 
\bea
H &= & - \frac{J}{2} \sum_{j=1}^L \left( a_j^{\dagger} a_{j+1} + a_{j+1}^{\dagger} a_j  \right) - \mu \sum_{j=1}^L \left( a_j^{\dagger} a_j -\frac{1}{2} \right) \nonumber\\
&& + \frac{1}{2} \sum_{j=1}^L \left( \Delta a_j^{\dagger} a_{j+1}^{\dagger} + \Delta^* a_{j+1} a_{j} \right).
\eea
Without loss of generality, we set $J =1$. As before, via a Fourier transform
\begin{equation}
\begin{pmatrix}
a_{j} \\  
a_{j}^{\dagger} \\
\end{pmatrix} 
= L^{-1/2} \sum_{k \in \text{BZ}} e^{i k j} 
\begin{pmatrix}
a_{k} \\
a_{-k}^{\dagger} \\  
\end{pmatrix} ,
\end{equation}
where momentum modes are $k_n = \frac{2\pi}{L}(n+1/2), n=-L/2, ..., L/2-1$, the Hamiltonian in position space turns into a free fermionic one in momentum space
\begin{equation}
H  = \frac{1}{2} \sum_{k \in \text{BZ}} {\Phi}_k^{\dagger} \vec{R}(k) \cdot \vec{\sigma} \Phi_{k}, \qquad 
\Phi_k = 
\begin{pmatrix}
a_{k} \\
a_{-k}^{\dagger} \\  
\end{pmatrix} ,
\end{equation}
where the vector of parameters now reads 
\begin{equation}
\vec{R}(k) =
\begin{pmatrix}
0 \\
|\Delta| \sin k \\
\mu + J \cos k \\
\end{pmatrix}, \qquad R(k) =  \left|\vec{R}(k) \right|.
\label{RkKt}
\end{equation}
Similarly to SSH model, the Hamiltonian can be separated into two parts
\begin{equation}
H = H_{M} + H_{CP},
\end{equation}
where 
\begin{eqnarray}
H_{M} &=& -  \sum_{k>0} \left(\mu  +J \cos k \right) \left(  a_k^{\dagger} a_k - a_{-k} a_{-k}^{\dagger}\right), \\
H_{CP} &=&   \sum_{k>0} i |\Delta| \sin k \left( a_k^{\dagger} a_{-k}^{\dagger} - a_{-k} a_k  \right).   \label{HamiCP}
\end{eqnarray}
Note that $H_{CP}$ is the interaction part that creates ``Cooper pairs".\\

The Hamiltonian can be further diagonalized 
\begin{equation}
H 
=  \sum_{k>0} R(k) \left[ -\eta_{-k}\eta_{-k}^{\dagger} + \eta_k^{\dagger} \eta_k  \right],
\end{equation}
where
\begin{equation}
\begin{pmatrix}
\eta_{-k}^{\dagger} \\
\eta_k \\
\end{pmatrix} =
\begin{pmatrix}
\vec{v}_+ & \vec{v}_-
\end{pmatrix}^{\dagger}(k)
\Phi_k 
\label{KtBT}
\end{equation}
and we followed the definition of $\vec{v}_{\pm}$ in \eqref{evpm}. \\
This way, the ground state of the model can be specified by the condition
\begin{equation}
\eta_k |\psi_{gs}\rangle = 0,\qquad \forall k \in \text{BZ}.
\end{equation}
If denote the Fock vacuum of the operators $\{ a_k \}$ to be $|0 \rangle$, the ground state can be written as
\begin{equation}
| \psi_{gs} \rangle  = \prod_{k>0} \left[  \left| \sin \frac{\phi_k}{2} \right| e^{ -i e^{i \operatorname{arg} \Delta} \cot \frac{\phi_k}{2} a_k^{\dagger} a_{-k}^{\dagger}} \right] |0 \rangle, 
\label{KtGS0}
\end{equation}
where we used the spherical coordinate to rewrite $\vec{R}(k)$
\begin{equation}
\cot \frac{\phi_k}{2} = \frac{| R_2 |}{R-R_3}, 
\label{phik}
\end{equation}
analogously to the SSH model.

Next, we introduce the $\operatorname{SU}(2)$ algebra generators $J^{(k)}_0$, $J^{(k)}_\pm$ for positive $k$ in terms of the fermion creation and annihilation operators  
\begin{eqnarray}
J_0^{(k)} &=& \frac{1}{2} \left( a_k^{\dagger} a_k - a_{-k} a_{-k}^{\dagger} \right), \\
J_+^{(k)} &=& a_k^{\dagger} a_{-k}^{\dagger}, \\
J_-^{(k)} &=& a_{-k}^{\dagger} a_{k},
\end{eqnarray}
such that the ground state \eqref{KtGS0} can be rewritten in terms of $\operatorname{SU}(2)$ coherent states
\begin{equation}
\begin{split}
| \psi_{gs} \rangle 
=& \prod_{k>0} \cos \frac{\pi-\phi_k}{2} \exp \left[   e^{-i \frac{\pi}{2}} \tan \frac{\pi-\phi_k}{2} J_+^{(k)} \right]   \\
& \qquad \qquad \qquad \bigotimes_{k \in \text{BZ}} \left| 1/2; -1/2 \right\rangle_k, \\
=& \prod_{k>0} \exp \left[ \frac{\pi-\phi_k}{2} e^{-i \frac{\pi}{2}} J_+^{(k)} -\frac{\pi-\phi_k}{2} e^{i \frac{\pi}{2}} J_-^{(k)} \right]  \\
& \qquad \qquad \qquad \bigotimes_{k \in \text{BZ}} \left| 1/2; -1/2 \right\rangle_k, \\
\end{split}
\end{equation}
where in the last line we have applied the BCH relation in \eqref{gnBCH}. Finally, the Hamiltonian of Kitaev model can be rewritten in terms of $\operatorname{SU}(2)$ algebra generators as
\begin{equation}
\begin{split}
H
=&   \sum_{k>0}\left[ -2 \left(\mu  +J \cos k \right) J_0^{(k)} \right. \\
& \qquad  \qquad \left. + i \Delta \sin k J_+^{(k)} - i \Delta^* \sin k J_-^{(k)}  \right] .
\end{split}
\end{equation}

\begin{figure}[h!]
\centering
\includegraphics[width =0.4\textwidth]{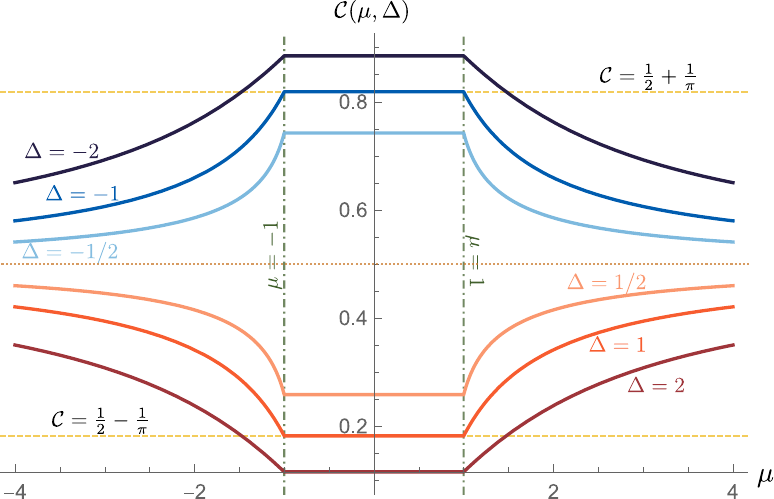}
\caption{Complexity $\mathcal{C}(s=1; \mu , \Delta)$ as a function of $\mu$ for various $\Delta$s ($\Delta = -2,-1,-1/2,1/2,1,2$). Between the two vertical gridlines $\mu = \pm 1$, spread complexity is $\mu$-independent. The two dashed horizontal gridlines are the analytical results of $\mathcal{C}(|\mu|<1, \Delta = \pm 1)$, respectively. For $|\mu| \to \infty$, the spread complexities of various $\Delta$s approach the dotted horizontal gridline $\mathcal{C} =1/2$.}
\label{Kt_cmpl_mu}
\end{figure}

In order to compute the spread complexity in this model we again naturally select the reference state to be the ground state of $H_{CP}$ (``Cooper pair vacuum") given in \eqref{HamiCP}, with arbitrary positive $\Delta$. In terms of $\operatorname{SU}(2)$ coherent states, it is not difficult to find that
\begin{equation}
|G_{CP} \rangle = \prod_{k>0} e^{ \frac{\pi}{4} e^{-i \frac{\pi}{2}} J_+^{(k)} - \frac{\pi}{4} e^{+i \frac{\pi}{2}} J_-^{(k)} } \bigotimes_{k \in \text{BZ}} \left| 1/2; -1/2 \right\rangle_k.
\end{equation}
Then for $| \psi_{gs} \rangle$ as our target state we have the circuit
\begin{equation}
| \psi_{gs} \rangle 
= \prod_{k>0} e^{ -i \left( \operatorname{sgn} \Delta \cdot \frac{\pi-\phi_k}{2} -\frac{\pi}{4} \right) \left( J_+^{(k)} + J_-^{(k)} \right) } |G_{CP} \rangle. 
\end{equation}
Following the same steps as for the SSH, we derive the total spread complexity in the continuum limit as
\bea
\mathcal{C} (s=1;  \mu, \Delta)& =& \frac{1}{\pi} \int_0^{\pi} dk ~ \sin^2 \left( \operatorname{sgn} \Delta \cdot \frac{\pi-\phi_k}{2} -\frac{\pi}{4} \right)  \nonumber\\
&=& \frac{1}{\pi} \int_0^{\pi} dk ~ \frac{1 - \operatorname{sgn} \Delta \cdot \sin  \phi_k }{2},
\eea
where the relation to the physical parameters is
\begin{equation}
\sin \phi_k = \frac{|R_2|}{R}= \frac{|\Delta| \sin k}{\sqrt{(\mu + \cos k)^2 + (|\Delta|\sin k)^2}}.
\end{equation}
The behaviour of spread complexity $\mathcal{C} (s=1; \mu,\Delta)$ as a function of $\mu$ is shown in Fig. \ref{Kt_cmpl_mu}. Clearly, similarly to the SSH, we find a constant plateau in the topological phase.\\

\begin{figure}[h!]
\centering
\includegraphics[width =0.4\textwidth]{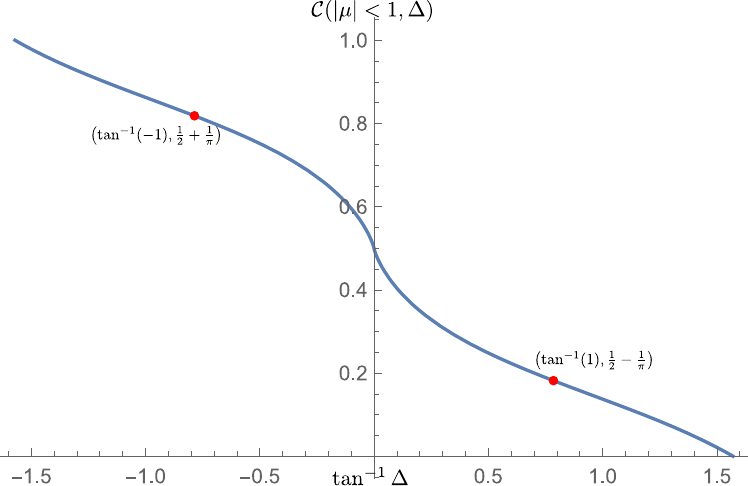}
\caption{Complexity $\mathcal{C}(s=1; |\mu|<1 , \Delta)$ as a function of $\Delta$. The two red dots are the analytical results of $\mathcal{C}(|\mu|<1, \Delta = \pm 1)$, respectively.}
\label{Kt_cmpl_Del}
\end{figure}

Namely, in the topological phase where $|\mu| <1$, we can derive the value of the spread complexity via integrating by parts
\begin{equation}
\begin{split}
 & \mathcal{C} (s=1;  |\mu|<1 , \Delta) \\
& \qquad = \frac{1 - \operatorname{sgn} \Delta}{2} + \frac{1}{\pi} \int_0^{\pi/2} \cos \phi_k \tan^{-1} \frac{\tan \phi_k}{\Delta} d \phi_k 
\end{split},
\end{equation}
since $\phi_k(k)$ is monotonic and inversely we also have
\begin{equation}
k = \varphi_k + \sin^{-1} \mu \sin \varphi_k, \quad
\tan \varphi_k = \frac{\tan \phi_k}{|\Delta|}.
\end{equation}
This implies that the spread complexity is independent of $\mu$ in the topological region. With some more effort we find this constant in terms of $\Delta$ 
\begin{equation}
\mathcal{C} (s=1;  |\mu|<1 , \Delta) 
= \frac{1}{2} - \frac{1}{\pi} \frac{\Delta \tan^{-1} \sqrt{\Delta^2-1}}{ \sqrt{\Delta^2 -1}} .
\end{equation}
This result is plotted on Fig. \ref{Kt_cmpl_Del}. Red dots are obtained by taking the $\Delta\to \pm 1$ limit in the formula above. In particular, the $\Delta \to 1$ limit reproduces the constant of the SSH model (see [41] for more discussion).

Note also that there is a discontinuity of the second derivative of spread complexity w.r.t $\Delta$. In particular,
\begin{equation}
\begin{split}
& \pi \frac{d^2}{d\Delta^2} \mathcal{C} (s=1; |\mu|<1 , \Delta) \\
& \qquad \qquad =\frac{2 \Delta ^2+1}{\Delta  \left(\Delta ^2-1\right)^2}-\frac{3 \Delta  \tan ^{-1}\left(\sqrt{\Delta ^2-1}\right)}{\left(\Delta ^2-1\right)^{5/2}},
\end{split}
\end{equation}
behaves as $\Delta^{-1}$ around $\Delta =0$.

\section{Appendix E: Relation to topological order parameters}\label{AppendixE}
In the main text, as well as in the additional example of the Kitaev chain, we showed that the spread complexity is clearly correlated with non-trivial topological phases of matter. A natural and very interesting question is then whether one can relate this quantity to any of the known topological order parameters or other commonly used ``witnesses" of topological phases? At the moment we do not have any definite answer to this interesting question and we only make a couple of comments below.

Firstly, it was noted in [28] that K-complexity based on Lie-algebra symmetry (including the $su(2)$ used here) can be seen as a Berry connection. In that work, it was also found that complexity corresponds to the volume in the Fubini-Study metric defined from the appropriately used coherent states. These notions seem similar in spirit to the probes employed e.g. in [38] such as e.g. the Zak's phase. Nevertheless, sharpening and better understanding of this link remains a future problem.

Secondly, also following [38] we tried to determine whether the spread complexity that we found in our paper can be written in terms of some expectation value of a position-space operator (maybe also in the spirit of position space entanglement). Below, using some guessing and/or reverse-engineering, we  show that the spread complexity of (formation of) the ground-state can be approximately related to the expectation value of  
\begin{equation}
\mathfrak{O} = \exp \left[ i \frac{2\pi}{L} \sum_j \left( i c_{Aj}^{\dagger} c_{Bj} + \text{h.c.} \right) \right],
\end{equation}
computed in the ground-state.

To see that, it is not difficult to show that 
for $\iota = A,B$ and $\epsilon \to 0+$,
\begin{eqnarray}
\left( \mathfrak{O}^{\dagger} \right)^{\epsilon} c_{\iota,j} \mathfrak{O}^{\epsilon} 
&=&   c_{\iota,i} + i \frac{2\pi}{L}\epsilon \sum_{j}  [ c_{\iota,i},  ic_{Aj}^{\dagger} c_{Bj} + \text{h.c.}  ]+ \mathcal{O}(\epsilon^2) \nonumber \\
&=& \sum_{\iota' = A,B} \left[ \delta_{\iota \iota'} - i \frac{2\pi}{L} \epsilon \sigma_{y,\iota\iota'} \right] c_{\iota',i}.
\end{eqnarray}
Thus
\begin{equation}
\mathfrak{O}^{\dagger} c_{\iota,i} \mathfrak{O}  = \sum_{\iota' = A,B} \left( e^{-i \frac{2\pi}{L} \sigma_y} \right)_{\iota \iota'}c_{\iota',i} .
\end{equation}

In the basis where the Hamiltonian is diagonalized,  one can find
\begin{eqnarray}
\mathfrak{O}^{\dagger} \chi_{\iota,k} \mathfrak{O}  &=& \left[ 
\begin{pmatrix}
\vec{v}_+^{\dagger} \\
 \vec{v}_-^{\dagger} \\
\end{pmatrix} e^{-i \frac{\pi}{4} \sigma_x} 
e^{-i \frac{2\pi}{L} \sigma_y}
e^{i \frac{\pi}{4} \sigma_x} 
\begin{pmatrix}
\vec{v}_+ & \vec{v}_- \\
\end{pmatrix}
 \right]_{\iota \iota'} \nonumber \\
&& \chi_{\iota',k } \nonumber \\
&=& \left[ 
\begin{pmatrix}
\vec{v}_+^{\dagger} \\
 \vec{v}_-^{\dagger} \\
\end{pmatrix}  
e^{i \frac{2\pi}{L} \sigma_z}
\begin{pmatrix}
\vec{v}_+ & \vec{v}_- \\
\end{pmatrix}
 \right]_{\iota \iota'} \chi_{\iota',k }.
\end{eqnarray}

The expectation value of $\mathfrak{O}$ in the ground-state $|\Omega \rangle$ is then
\begin{eqnarray}
&& \langle \Omega | \mathfrak{O} | \Omega \rangle  = \langle 0 | \mathfrak{O} \prod_{k \in \text{BZ}}\left( \mathfrak{O}^{\dagger}\chi_{-,k} \mathfrak{O} \right) \prod_{q \in \text{BZ}} \chi_{-,q}^{\dagger} |0 \rangle \nonumber \\
&=&  \prod_{k \in \text{BZ}} \left[ 
\begin{pmatrix}
\vec{v}_+^{\dagger}(k) \\
 \vec{v}_-^{\dagger}(k) \\
\end{pmatrix}  
e^{i \frac{2\pi}{L} \sigma_z}
\begin{pmatrix}
\vec{v}_+(k) & \vec{v}_-(k) \\
\end{pmatrix}
 \right]_{-,-}  \nonumber \\
&\approx&  \prod_{k \in \text{BZ}} \left[ 1 + i \frac{2\pi}{L} \vec{v}_-^*(k) \cdot \sigma_z \vec{v}_-(k) + \mathcal{O}(L^{-2}) \right]
\end{eqnarray}
for large $L$. 
Hence it could be rewritten into
\begin{equation}
 \langle \Omega | \mathfrak{O} | \Omega \rangle  
=  \exp \left[  i \frac{2\pi}{L} \sum_k \vec{v}_-^*(k) \cdot \sigma_z \vec{v}_-(k) + \mathcal{O}(L^{-2}) \right].
\end{equation}

On the other hand,  given that in spherical coordinates system
\begin{equation}
\vec{v}_- = 
\begin{pmatrix}
\sin \frac{\phi_k}{2}e^{-i \psi_k} \\
- \cos \frac{\phi_k}{2} \\
\end{pmatrix},
\end{equation}
one can find
\begin{equation}
 \vec{v}_-^*(k) \cdot \sigma_z \vec{v}_-(k) = -  \cos \phi_k = 2 \mathcal{C}_k -1.
\end{equation}

This way we have an approximate relation in the SSH model
\begin{eqnarray}
 \langle \Omega | \mathfrak{O} | \Omega \rangle  
&=&  \exp \left[  i \frac{2\pi}{L} \sum_k \left(2\mathcal{C}_k - 1\right) + \mathcal{O}(L^{-2}) \right] \nonumber \\
&=& \exp \left[  i 4 \pi \mathcal{C}(t_1,t_2) + \mathcal{O}(L^{-2}) \right].
\end{eqnarray}

In order to get some more physical intuition behind the expectation value of $\mathfrak{O}$, we consider its exponent that is a sum of operators:
\begin{equation}
\mathfrak{P}_j = i c_{Aj}^{\dagger} c_{Bj} + \text{h.c.} .
\end{equation}
In the large $L$ limit,  the sum over $\langle \Omega | \mathfrak{P}_j | \Omega \rangle$ over all $j$s is $2L$ times the spread complexity (density). Below we consider two examples to show that $\langle \Omega | \mathfrak{P}_j | \Omega \rangle$ is related to the entanglement between site $j$ and its complement in position space. For convenience, we use the notation of first quantization.  The Hamiltonian can be rewritten in this language as
\begin{eqnarray}
\hat{H} &=& t_1 \sum_j^L \left( |j,B \rangle \langle j,A | + \text{h.c.} \right) \nonumber  \\
 && - t_2 \sum_j^L \left( |j+1,A \rangle \langle j,B | + \text{h.c.} \right).
\end{eqnarray}
In this notation, operator $\mathfrak{P}_j$ has Pauli matrix $\sigma^y$ representation. In particular,
\begin{equation}
\mathfrak{P}_j = i | j, A \rangle \langle j, B | + \text{h.c.}.
\end{equation}

Let us now consider the expectation value of $\mathfrak{P}_j$ for the two different values of parameters $(t_1,t_2)$ that correspond to trivial as well as topological phases:

\begin{enumerate}
\item $t_1 =1,  t_2=0$ (Non-topological phase)

In this case
\begin{equation}
\hat{H} \left( | j, A \rangle \pm | j, B \rangle  \right) = \pm  \left( | j, A \rangle \pm | j, B \rangle  \right)
\end{equation}
are the two eigenstates on site $j$. Thus the ground-state of the model can be written as
\begin{equation}
|\Omega^{NP} \rangle = \bigotimes_{j=1}^L \frac{1}{\sqrt{2}}\left( | j, A \rangle - | j, B \rangle  \right).
\end{equation}
The expectation value of $\mathfrak{P}_j$ can then be re-written as a trace over all sites but $j$:
\begin{equation}
\langle \Omega^{NP} | \mathfrak{P}_j | \Omega^{NP} \rangle = \operatorname{Tr}_j \mathfrak{P}_j \rho^{NP}_j,
\end{equation}
where
\begin{equation}
\rho^{NP}_j = \operatorname{Tr}_{\neg j} | \Omega^{NP} \rangle \langle \Omega^{NP} |.
\end{equation}
It is not difficult to find
\begin{equation}
\rho^{NP}_j =\frac{1}{2} \left( | j, A \rangle - | j, B \rangle  \right) \left( \langle j, A | - \langle j, B |  \right).
\end{equation}
This is a density matrix of pure state. As a result its entanglement entropy is trivial and the expectation value of $\mathfrak{P}_j$ is correlated with trivial entanglement structure in this phase.

\item $t_1=0, t_2=-1$ (Topological phase)

In this case
\begin{equation}
\hat{H} \left( | j+1, A \rangle \pm | j, B \rangle  \right) = \pm  \left( | j+1, A \rangle \pm | j, B \rangle  \right)
\end{equation}
are the two eigenstates on site $j$. Thus the ground-state can now be presented as
\begin{equation}
|\Omega^{TP} \rangle = \bigotimes_{j=1}^{L-1} \frac{1}{\sqrt{2}}\left( | j+1, A \rangle - | j, B \rangle  \right).
\end{equation}
The expectation value of $\mathfrak{P}_j$ can be re-written by trace of all sites but $j$:
\begin{equation}
\langle \Omega^{TP} | \mathfrak{P}_j | \Omega^{TP} \rangle = \operatorname{Tr}_j \mathfrak{P}_j \rho^{TP}_j,
\end{equation}
where
\begin{equation}
\rho^{TP}_j = \operatorname{Tr}_{\neg j} | \Omega^{TP} \rangle \langle \Omega^{TP} |.
\end{equation}\\

It is not difficult to find that now
\begin{eqnarray}
\rho^{TP}_j &=&\frac{1}{4} \left( 1 + | j, A \rangle \langle j, A | + | j, B \rangle \langle j, B | \right.  \nonumber \\
&& \left.+ | j, A \rangle \langle j, A | \otimes | j, B \rangle \langle j, B | \right),  \nonumber \\
&=& \frac{1}{4} \left( \rho_{j,0}^{TP} \oplus 2 \rho_{j,1}^{TP} \oplus \rho_{j,2}^{TP}  \right).
\end{eqnarray}
This reduced density matrix consists of three sectors.  Among them, only the sector $\rho_{j,1}^{TP}$ can be expanded by $\sigma_j^x$,  $\sigma_j^y (= \mathfrak{P}_j )$, and $\sigma_j^z$. This implies that we can only retrieve part of "information" of the ground state via $\mathfrak{P}_j$ and therefore $\mathfrak{O}$, which leads to the difference from the non-topological phase. On the other hand, from the view of entanglement, the entanglement between site $j$ and its neighbours is nontrivial. 

\end{enumerate}

Concluding this part, it is too early to make a definite statement whether the spread complexity applied to the somewhat simpler and integrable models can be expressed in terms of the known probes of the topological phases. By the previous works as well as above intuitive arguments it is likely that it may be correlated with entanglement-like probes but the precise details should be spelled out. What we can say for sure is that, in more general and chaotic settings, spread complexity is a new measure sensitive to the spectrum of the model. We hope that further studies of this quantity will deepen our understanding of this tool and help to find appropriate spot for it also in the condensed matter toolkit.

\end{document}